\documentstyle[11pt]{article}

\def\texpsfig#1#2#3{\vbox{\kern #3 \hbox{\includegraphics{#1}\kern #2}}
\typeout{(#1)}}
\input epsf

\begin{document}

\begin{center}
{\bf \huge Energy Estimation of UHE Cosmic Rays using the 
Atmospheric Fluorescence Technique}

\vspace{0.2in}

{\bf C. Song$^3$, Z. Cao$^4$, B.R. Dawson$^1$, B.E. Fick$^2$,
P. Sokolsky$^4$, X. Zhang$^3$}

{\it $^1$ University of Adelaide, Adelaide, South Australia 5005, Australia}

{\it $^2$ Enrico Fermi Institute, University of Chicago, Chicago, IL 60637, USA}

{\it $^3$ Department of Physics, Columbia University, New York, NY 10027, USA}

{\it $^4$ High Energy Astrophysics Institute, 

University of Utah, Salt Lake City, UT 84112, USA}

\end{center}

\vspace{0.5in}

\begin{center}
{\bf \large Abstract}
\end{center}
\vspace{-0.5ex}

We use the CORSIKA air shower simulation program to review the
method for assigning energies to ultra-high energy (UHE) cosmic
rays viewed with the air fluorescence technique.  This technique
uses the atmosphere as a calorimeter, and we determine the
corrections that must be made to the calorimetric energy to yield
the primary cosmic ray energy.

\vspace{0.3in}



e-mail address : csong@cosmic.utah.edu

Fax number : 1-801-581-6256

Phone number : 1-801-581-4306

\newpage

\section{Introduction}

One of the goals in a cosmic ray detection experiment is to
determine the energy of the incident particles.  Unfortunately,
the primary energy cannot be measured directly at high energies
where the flux is very low. Instead we take advantage of the
cascades (or extensive air showers) produced by the cosmic rays
in the atmosphere.  The secondary particles that make up the
cascade can be detected at ground level, as can the Cerenkov
light that they produce.  Alternatively, one can detect the
atmospheric nitrogen fluorescence light induced by the passage of
the shower.  This technique, employed by the Fly's Eye detector
and its successor the High Resolution Fly's Eye (HiRes) has the
advantage that one can measure the number of charged particles in
the shower as a function of depth in the atmosphere, $N_{ch}(X)$,
where $X$ is measured in $g/cm^2$.  We treat the atmosphere as a
calorimeter, and the primary cosmic ray energy can be estimated
by integrating the longitudinal profile $N_{ch}(X)$ and making
corrections for ``unseen'' energy.

In the past, the energy of a pure electromagnetic shower has been
determined by \cite{rossi},

\begin{eqnarray}
E_{em} & = & {E_{c} \over X_{0}} \int_{0}^\infty N_{e}(X)dX,
\end{eqnarray}
where $X_{0}$ is the electron radiation length in air, $E_{c}$ is
the critical energy of an electron in air, and $N_{e}$ is the
number of electrons in the shower.  This equation implies that
the electromagnetic energy is the total track length of all
charged particles multiplied by an energy loss rate $dE/dX$ given
by $E_{c}/X_{0}$.  One source of error here is the numerical
value of the critical energy $E_c$, which has two definitions
attributed to Rossi \cite{data}, and Berger and Selter
\cite{table}.  We turn to simulations to check the result and
avoid this confusion.

We do this using the modern shower simulation package CORSIKA
\cite{corsika}.  We simulate ultra-high energy showers complete
with realistic fluctuations and realistic distributions of the
energies of shower particles.  An example is shown in Figure~1
where we plot the energy spectra of shower particles at the depth
of the maximum size.  Particle energies cover a wide range and lose 
energy to ionization at different rates (see inset to figure).
 
For these simulations we replace Eq.(1) with a more general
expression for the calorimetric energy, 

\begin{eqnarray}
E_{cal} & = & \alpha \int_{0}^\infty N_{ch}(X)dX,
\end{eqnarray}
where we integrate the charged particle longitudinal profile.  We
replace the constant in Eq.(1) with a parameter $\alpha$
representing the mean ionization loss rate over the entire
shower.  (This factor will be approximately equal to $E_c/X_0$
and is calculated below).  Given $E_{cal}$ we must then make a
correction to determine the cosmic ray energy $E_0$.  The
correction takes account of energy carried by high energy muons
and neutrinos that ultimately deposit most of their energy in the
ground.  It also takes account of the small amount of energy that
is lost to nuclear excitation.  This ``missing'' energy has
previously been parametrized by Linsley \cite{linsley} and the 
Fly's Eye group \cite{balt}.

In this paper we first describe some characteristics of the
CORSIKA shower simulation package.  We then use CORSIKA to
simulate gamma-ray induced air showers (which have a very small
``missing energy'' component) to check the calorimetric energy
method.  Finally we simulate proton and iron induced showers to
calculate the ``missing energy'' corrections for primary energies
up to $10^{20}$eV.

\section{The Simulations}

CORSIKA is a versatile package for simulating air showers over a
wide range of primary energies.  Choices are available for the
hadronic interaction model at the higest energies, and we have
chosen the QGSJET \cite{qgsjet} description which is in good agreement
with Fly's Eye measurements.  Within CORSIKA electromagnetic
sub-showers are simulated with the EGS4 code. In EGS4, the
cross sections and branching ratios are extended to $10^{20}$eV
with the assumption that QED is valid to these energies.
In order to reduce CPU
time, a thinning algorithm was selected within CORSIKA. That is,
if the total energy of secondary particles from a given
interaction falls below $10^{-5}$ of the primary energy, only one
of the secondaries is followed, selected at random according to
its energy $E_{i}$ with a probability of $p_{i} = E_{i}/\sum_{j}
E_{j}$.  The sum does not include neutrinos or particles with
energies below the preset thresholds.  In our simulations the
threshold energies are 300, 700, 0.1 and 0.1 MeV for hadrons,
muons, electrons and photons respectively.  Particles below the
threshold energies are not followed by the simulation.  We chose
an observation level 300 m above sea level and we simulated showers 
with zenith angles of 45 degrees.

\section{Calorimetric energy of an air shower}

Consider a purely electromagnetic shower.  The primary particle
energy $E_{em}$ can be approximated by

\begin{eqnarray}
E_{em} & \cong & \int_{\epsilon}^\infty \triangle E(k) {\cal N}_{e}(k) dk,
\end{eqnarray}
where ${\cal N}_{e}(k)$ is the differential energy spectrum  
of electrons with kinetic
energy $k$ and $\triangle E(k)$ is the energy loss by each of
those electrons in the calorimeter via ionization.  This is only
an approximation because we have only included particles with
kinetic energies above a threshold $\epsilon$.  This is
consistent with our simulations where we must impose a threshold
of 0.1\,MeV for photons, electrons and positrons.  This integral
can be carried out by summing over all the electrons produced in
the simulation.

We rearrange Eq.(3) and include the energy spectrum of particles
as a function of atmospheric depth,

\begin{eqnarray}
{\cal N}_{e}(k) & = & \int_{0}^\infty N_{e}(X) n_{e}(k,X) 
{dX \over \triangle X(k),}
\end{eqnarray}
where $\triangle X(k)$ is the mean free path of electrons as a
function of $k$, $N_{e}(X)$ is the total number of electrons at
depth $X$, and $n_{e}(k,X)$ is the normalized electron energy
spectrum. Then, the electromagnetic energy is approximated by

\begin{eqnarray}
E_{em} & \cong & \int_{0}^\infty N_{e}(X) \left(
\int_{\epsilon}^\infty { {\triangle E} \over {\triangle X} } (k)
n_{e} (k,X) dk \right ) dX
\end{eqnarray}

The {\it age} parameter rather than the depth is often used to
describe the stage of development of a shower.  The energy
spectrum of electrons can then be parametrized in terms of age.
Since the age parameter is really only valid for a pure
electromagnetic cascade, and since we will use the parameter in
reference to hadronic showers, we will refer to our parameter as
the {\it pseudo age}.  We define it as

\begin{equation}
S(X) = { { 3 \cdot (X - X_{1}) } \over 
{ (X - X_{1}) + 2 \cdot ( X_{max} - X_{1} ) } }
\end{equation}
where $X_{1}$ is the depth of first interaction and $X_{max}$ is
the depth at which the shower reaches maximum size.  Under this
definition $S(X_{1}) = 0$, $S(X_{max}) = 1$ and $S(\infty) = 3$.

One can then calculate the mean ionization loss rate ($dE/dX$)
for the electrons in the shower (with energies $>\epsilon$) at
age $S$,

\begin{eqnarray}
\alpha(S) & = & \int_{\epsilon}^\infty { {\triangle E} \over {\triangle X} } 
(k) \widetilde{n}_{e} (k,S) dk
\end{eqnarray}
where $\widetilde{n}_e$ is now a function of $S$.  For comparison
with Eq.(1) we rewrite Eq.(5) as:

\begin{eqnarray}
E_{em} & \cong & <\alpha>_{S} \int_{0}^\infty N_{e}(X) dX,
\end{eqnarray}
where

\begin{eqnarray}
<\alpha>_{S} & = & {{\sum_{i} <N_{e}>_{\triangle S_{i}} \cdot 
\alpha(S)_{\triangle S_{i}} }
\over { \sum_{i} <N_{e}>_{\triangle S_{i}} } },
\end{eqnarray}
and $<N_{e}>_{\triangle S_{i}}$ is the average number of electrons
within a {\it pseudo age} bin $\triangle S_{i}$.

We have simulated $10^{17}$eV showers initiated by photons,
protons and iron nuclei in order to calculate the mean energy
loss rate over the entire shower, $<\alpha>_{S}$.  We use bins of
$\triangle S_{i} =0.1$.  Figure 2 shows $\alpha(S)$ as a function
af age and we find $<\alpha>_{S}$ is 2.186, 2.193 and 2.189
MeV/(g/cm$^{2}$) for gamma, proton and iron induced showers
respectively. All the errors in those $<\alpha>_{S}$s are less than
0.1\%.

This compares with the value of the ratio $E_c/X_0=2.18$
MeV/(g/cm$^{2}$) used by the Fly's Eye analysis, where the values
were taken to be $E_c=81$MeV and $X_0$=37.1 g/cm$^{2}$ \cite{radl}.
This agreement may be a coincidence, since more recent values of the
parameters from \cite{data} are $E_c=86$MeV (using Rossi's
definition) and $X_0$=36.7 g/cm$^{2}$, giving a ratio which is
7\% higher than the typical simulation value of $<\alpha>_{S}$.
However, we note that the simulation results only include the
energy loss rates for particles above the 0.1\,MeV threshold.

Figure 3 shows average shower profiles as a function of age for
different primary masses and energies, with the shower size
normalized to 1 at $S = 1$. The difference in the average
proton-induced shower profile at three different primary energies
is smaller than the difference between the proton and iron
average profiles at one energy. In other words, the shape of the
shower development curve as a function of $S$ is quite
independent of primary energy or primary mass. It is also well
known that for photon primaries the energy spectrum of the shower 
particles is a function of $S$ only. We found it is also true for 
hadronic showers in our Monte Carlo study. Hence we can assume that 
our result for $<\alpha>_{S}$ can be applied over a range of 
primary masses and energies.

We now apply Eq.(2) to some gamma-ray initiated CORSIKA showers,
with $\alpha = 2.19$MeV/(g/cm$^{2}$).  The CORSIKA energy
thresholds are set as described above with a threshold energy of
0.1\,MeV for photons, electrons and positrons.  We integrate the
shower development curves in two ways for comparison.  In the
first we numerically integrate the CORSIKA output which bins the
development curve in 5 g/cm$^2$ increments.  Alternatively we
fit a Gaisser-Hillas function (with variable $X_0$,
$X_{max}$, $N_{max}$ and $\lambda$) to the CORSIKA output then
integrate the function.  Both methods give results which agree 
at a level of better than 1\%.

Table 1 shows the results for 500 showers.  The calorimetric
energy is about 10\% lower than the true value.  This is true
even when we switch off processes which are not purely
electromagnetic, namely $\mu^+\mu^-$ pair production and
photo-nuclear reactions, which have small but important
cross-sections in gamma-ray initiated showers.  Two hundred such
showers were generated and the results are also shown in Table 1,
where we see that these showers have no muon content as expected.
However the deficit in the calorimetric energy remains close to
10\%.

The solution to the problem is related to the simulation energy
threshold of 0.1\,MeV.  We have made a detailed study with
CORSIKA of the energy loss mechanisms and the characteristics of
particles around 0.1\,MeV.  In particular we have summed
the energy of particles that drop below the 0.1\,MeV threshold.
Table 2 shows that at $10^{17}$eV, 88.4\% of the primary energy
is lost to the atmosphere through ionization by particles above
0.1\,MeV.  Electrons in the shower with energies below 0.1\,MeV
carry 9.0\% of the primary energy, while sub-0.1\,MeV photons
carry 1.2\% of the primary energy.  And the calorimetric energy
derived by Eq.(2) is 88.8\% of the primary energy, a good match
to the ionization energy loss by particles above 0.1\,MeV.

We assume that the sub-0.1\,MeV particles will eventually lose
energy to ionization.  The nitrogen fluorescence efficiency is
proportional to the ionization loss rate, so experiments like
HiRes will detect light in proportion to the energy loss, even
for very low energy particles.  Thus the problem we experience
with reconstructing the energy of CORSIKA simulations will not
occur with the real shower data.  So for the further CORSIKA
studies described below we have added 10\% of the primary energy
to the integrated energy loss result (from Eq.(2)) to take
account of the sub-0.1\,MeV particles that do not appear in the
CORSIKA output.

\section{Energy estimation for hadronic showers}

We have described the calorimetric energy estimation for
gamma-ray induced showers.  We next consider hadronic showers
where we expect the calorimetric energy to fall short of the
primary energy because of so called ``missing energy'' - that
energy channeled into neutrinos, high energy muons, and nuclear
excitation.  Much of this energy is deposited into the ground and
is not visible in the atmospheric calorimeter.  The first
estimate of missing energy was obtained by Linsley \cite{linsley}
who made measurements of electron and muon sizes at ground level
and assessed the energy content of these components.  The Fly's
Eyes group parametrized Linsley's estimates as \cite{{balt}}:
\begin{equation}
{E_{cal} / E_{0}} = 0.990 - 0.0782 \cdot E_{0} ^ {-0.175}
\end{equation}
where $E_{0}$ is the primary energy and $E_{cal}$ is the
calorimetric energy derived from Eq.(2), both in units of
$10^{18}$eV. This parametrization was said to be valid for
$10^{15}$eV$< E_{0} < 10^{20}$eV.

We have simulated proton and iron initiated showers at 8 primary
energies from $3\times10^{16}$eV to $10^{20}$eV using CORSIKA.
We apply Eq.(2) (with a mean energy loss rate of
2.19\,MeV/(g/cm$^2$)) by fitting a Gaisser-Hillas profile to the
CORSIKA development curve and extrapolating the profile to infinity
and then integrating the function.  We
then add 10\% of the primary energy to this result to take
account of the CORSIKA threshold effect.  Finally we compare this
calorimetric energy with the primary energy, as shown in
Figure~4. It is physically reasonable that the missing energy 
should decrease with increasing primary energy.  Because of 
relativistic effects, charged pions produced in more energetic 
showers have an increased chance of interacting rather than 
suffering decay, reducing the fraction of energy immediately 
directed into muon and neutrino production. For comparison 
we also show Linsley's result.  The solid line in Figure 4 
shows the average behavior for proton and iron showers, 
which we express here as a function of $E_{cal}$
(for practical convenience) in units of $10^{18}$eV,

\begin{equation}
{E_{cal} / E_{0}} = (0.959 \pm 0.003) - (0.082 \pm 0.003) \cdot E_{cal} ^ {-(0.150 \pm 0.006)}
\end{equation}
which is valid for $3\times10^{16}$eV$< E_{0} < 10^{20}$eV.
Unfortunately it is never possible to know the primary particle
mass on a shower-by-shower basis, so this average correction must
be used.  This lack of knowledge translates into an energy
uncertainty, which is at most about 5\% if the primary is
hadronic.  If the primary particle is a gamma-ray this assumption
will overestimate the energy by up to 20\%.  Of course, if the
shower development profile is obviously anomalous (as expected
for gamma-ray showers above $10^{19}$eV due to the LPM effect)
gamma-ray primaries can be recognized and this systematic can be
avoided.

In an experiment like HiRes, atmospheric nitrogen fluorescence
provides a measurement of ionization energy deposition, since the
yield of fluorescence photons is proportional to this energy
deposition \cite{fluo}.  In the reconstruction process we convert
the amount of light emitted by the shower at a particular depth
to a number of charged particles, assuming that those charged
particles are ionizing at the mean ionization rate which is a 
function of temperature and density \cite{fluo}. This is taken
into account in our analysis of real showers when we calculate 
the number of ionizing particles at a particular atmospheric 
depth. We then perform the path length integral (Eq.(2)), 
multiply by the mean ionization loss rate of 2.19\,MeV/(g/cm$^2$)) 
and then make a correction for missing energy (Eq.11).

\section{Conclusion}

We have re-investigated the veracity of estimating cosmic ray
energy by using the atmosphere as a calorimeter.  We have
determined that, provided we use an appropriate mean energy loss
rate, the technique provides a good estimate of primary energy
for gamma-ray induced showers.  For hadronic showers we have
derived a correction function which accounts for energy not
deposited in the atmosphere, so that the technique also returns a
good estimate of primary energy for these showers.

\section{Acknowledgments}

We are grateful to Dieter Heck for valuable answers to our
questions about CORSIKA.  The support from the Center for High
Performance Computing at the University of Utah is gratefully
acknowledged.

\newpage

\footnotesize
\begin{table}
\begin{center}
\begin{tabular}{|c||c|c|c||c|c|c|} \hline
& \multicolumn{3}{|c||}{With $\mu^+ \mu^-$ \& $\gamma N$}
& \multicolumn{3}{|c|}{Without $\mu^+ \mu^-$ \& $\gamma N$} \\ \hline\hline
$E_{0}$, eV  & $E_{cal}/E_{0}$ &    $N_{\mu}$    &   $N_{max}$ 
 & $E_{cal}/E_{0}$ &    $N_{\mu}$    &      $N_{max}$ 
\\ \hline
$10^{16}$&  0.888 $\pm$ 0.004 & (2.146 $\pm$ 0.795)$10^3$
& (8.324 $\pm$ 0.392)$10^6$ 
&  0.897 $\pm$ 0.003 & 0.000 $\pm$ 0.000 & (8.448 $\pm$ 0.413)$10^6$
\\ \hline
$10^{17}$&  0.888 $\pm$ 0.005 & (2.823 $\pm$ 1.369)$10^4$
& (7.881 $\pm$ 0.339)$10^7$ 
&  0.898 $\pm$ 0.004 & 0.000 $\pm$ 0.000 & (7.967 $\pm$ 0.392)$10^7$
\\ \hline
$10^{18}$&  0.889 $\pm$ 0.004 & (3.185 $\pm$ 0.916)$10^5$ 
& (7.439 $\pm$ 0.271)$10^8$ 
&  0.898 $\pm$ 0.003 & 0.000 $\pm$ 0.000 & (7.558 $\pm$ 0.281)$10^8$
\\ \hline
\end{tabular}
\caption{Results of CORSIKA simulations of gamma-ray induced air showers
at three primary energies.  The right-hand half of the table
shows results from simulations where photo-nuclear and muon pair
production processes have been switched off.  The uncertainties
shown are root mean squared errors.}
\label{tab:swon}
\end{center}
\end{table}

\newpage

\normalsize
\begin{table}
\begin{center}
\begin{tabular}{|c||c|c|c||c|} \hline
$E_{0}$, eV     &  $E_{loss}/E_{0}$  &  $E_{e}(<$0.1 MeV)$/E_{0}$  
&  $E_{\gamma}(<$0.1 MeV)$/E_{0}$  &  $E_{cal}/E_{0}$   \\ \hline
    $10^{16}$   & 0.888 $\pm$ 0.003  & 0.090 $\pm$ 0.001
& 0.010 $\pm$ 0.001   & 0.888 $\pm$ 0.004   \\ \hline
    $10^{17}$   & 0.884 $\pm$ 0.005  & 0.090 $\pm$ 0.001 
& 0.012 $\pm$ 0.003   & 0.888 $\pm$ 0.005   \\ \hline
    $10^{18}$   & 0.876 $\pm$ 0.007  & 0.092 $\pm$ 0.002
& 0.018 $\pm$ 0.005   & 0.889 $\pm$ 0.004   \\ \hline
\end{tabular}
\caption{Results from a study of energy conservation within CORSIKA.  Gamma-ray
showers were simulated at three primary energies $E_0$.  $E_{loss}$
refers to the energy lost to the atmosphere through ionization by
charged particles with energies above 0.1\,MeV.  The fraction of
the primary energy carried by sub-0.1\,$MeV$ electrons and
photons is shown in the next two columns.  The fraction of
primary energy determined by the calorimetric equation (final
column) is consistent with $E_{loss}/E_0$.  Again, all
uncertainties are r.m.s.}
\label{tab:gamma}
\end{center}
\end{table}

\newpage

\begin{figure}
\begin{center}
\leavevmode
\hbox{%
\epsfxsize=6.0in
\epsffile{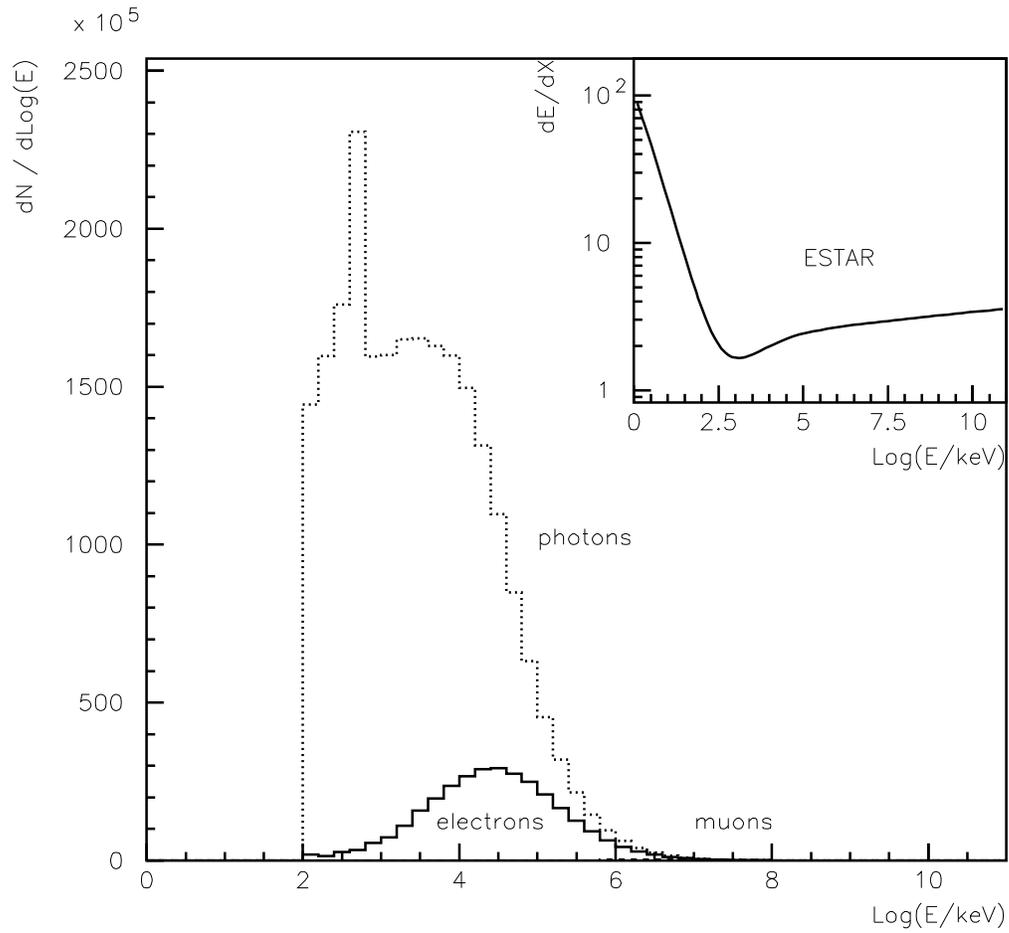}}
\end{center}
\caption{The mean energy spectra of photons, electrons and muons
at $S = 1$ for 200 proton showers at $10^{17}$ eV.
The spike in the photon spectrum corresponds to electron-positron 
annihilation. The inset shows the energy loss rate (in MeV/g/cm$^2$) by
ionization of electrons in dry air over the same energy range as
the main figure. The ESTAR code produced by US National Institute
of Science and Technology (NIST) was used below 10 GeV\cite{estar}
and this curve is extrapolated into the region above 10 GeV.}
\label{fig.1}
\end{figure}

\newpage

\begin{figure}
\begin{center}
\leavevmode
\hbox{%
\epsfxsize=6.0in
\epsffile{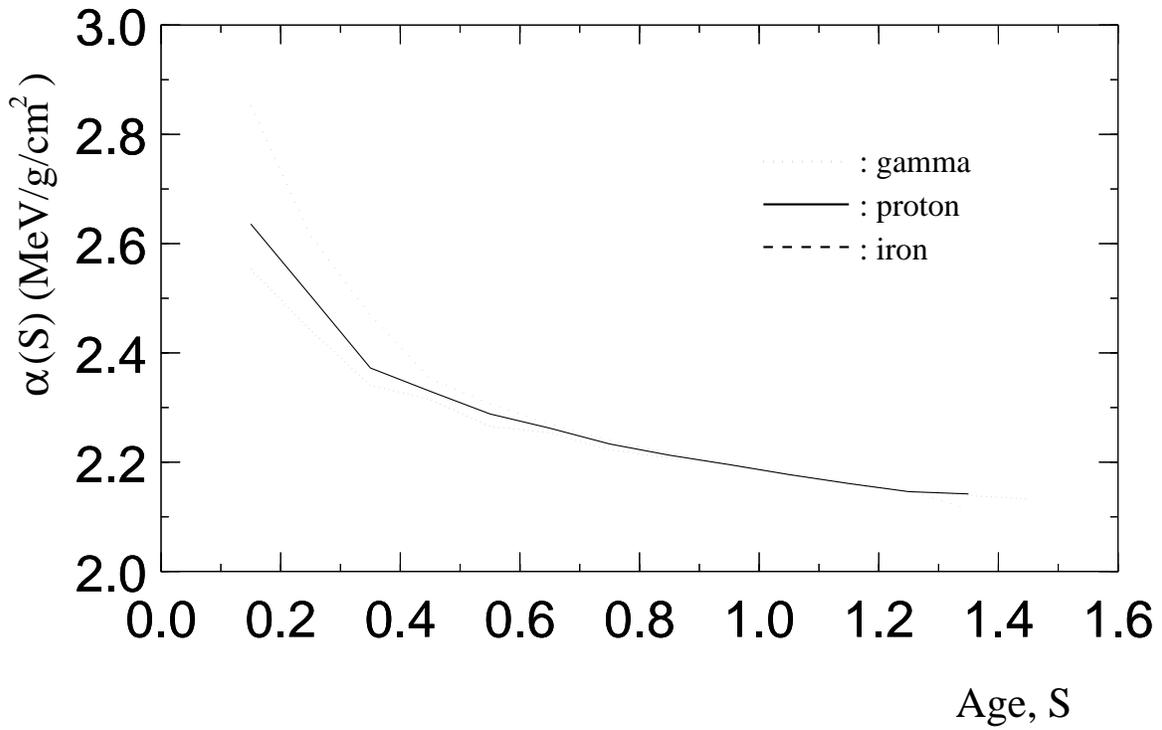}}
\end{center}
\caption{The mean ionization loss rate $dE/dX$ as function of S for
gamma-ray, proton, and iron induced showers at $10^{17}$eV.}
\label{fig.2}
\end{figure}

\newpage

\begin{figure}
\begin{center}
\leavevmode
\hbox{%
\epsfxsize=6.0in
\epsffile{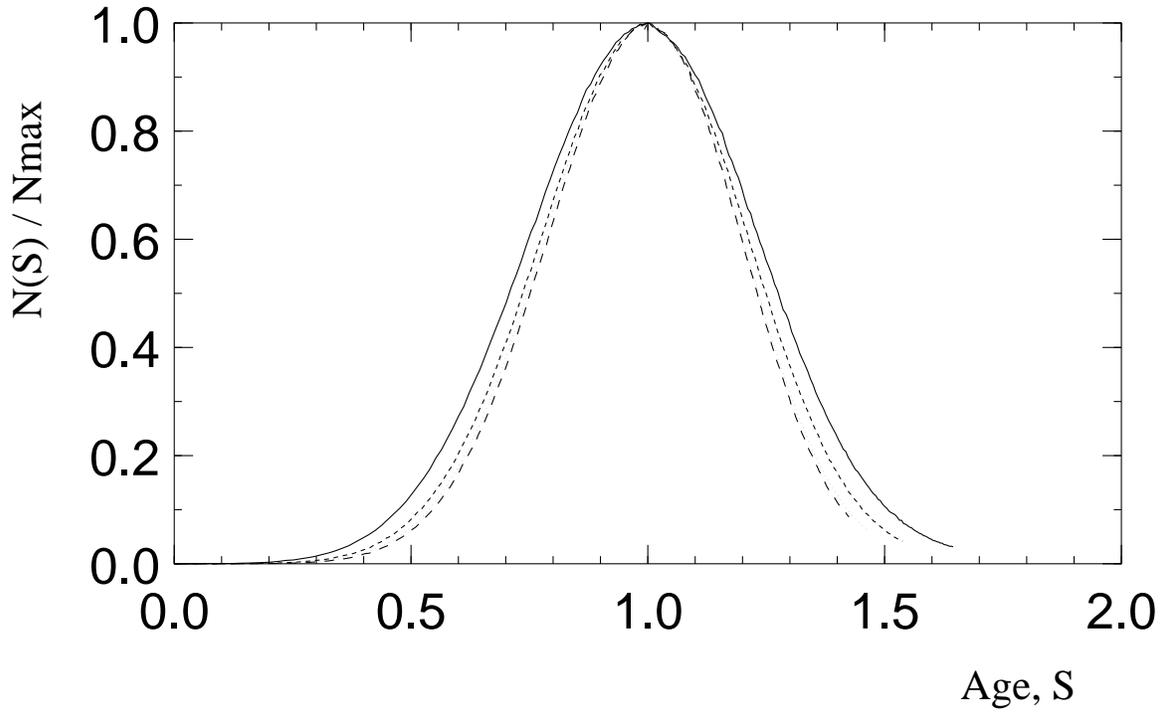}}
\end{center}
\caption{The solid line is an  average shower profile for 200 iron 
induced showers at $10^{17}$ eV. The other three lines are
average shower profiles for 200 proton showers at 10$^{17}$eV
(short dashed line), 10$^{18}$eV (dotted line) and 10$^{19}$eV
(long dashed line).}
\label{fig.3}
\end{figure}

\newpage

\begin{figure}
\begin{center}
\leavevmode
\hbox{%
\epsfxsize=6.0in
\epsffile{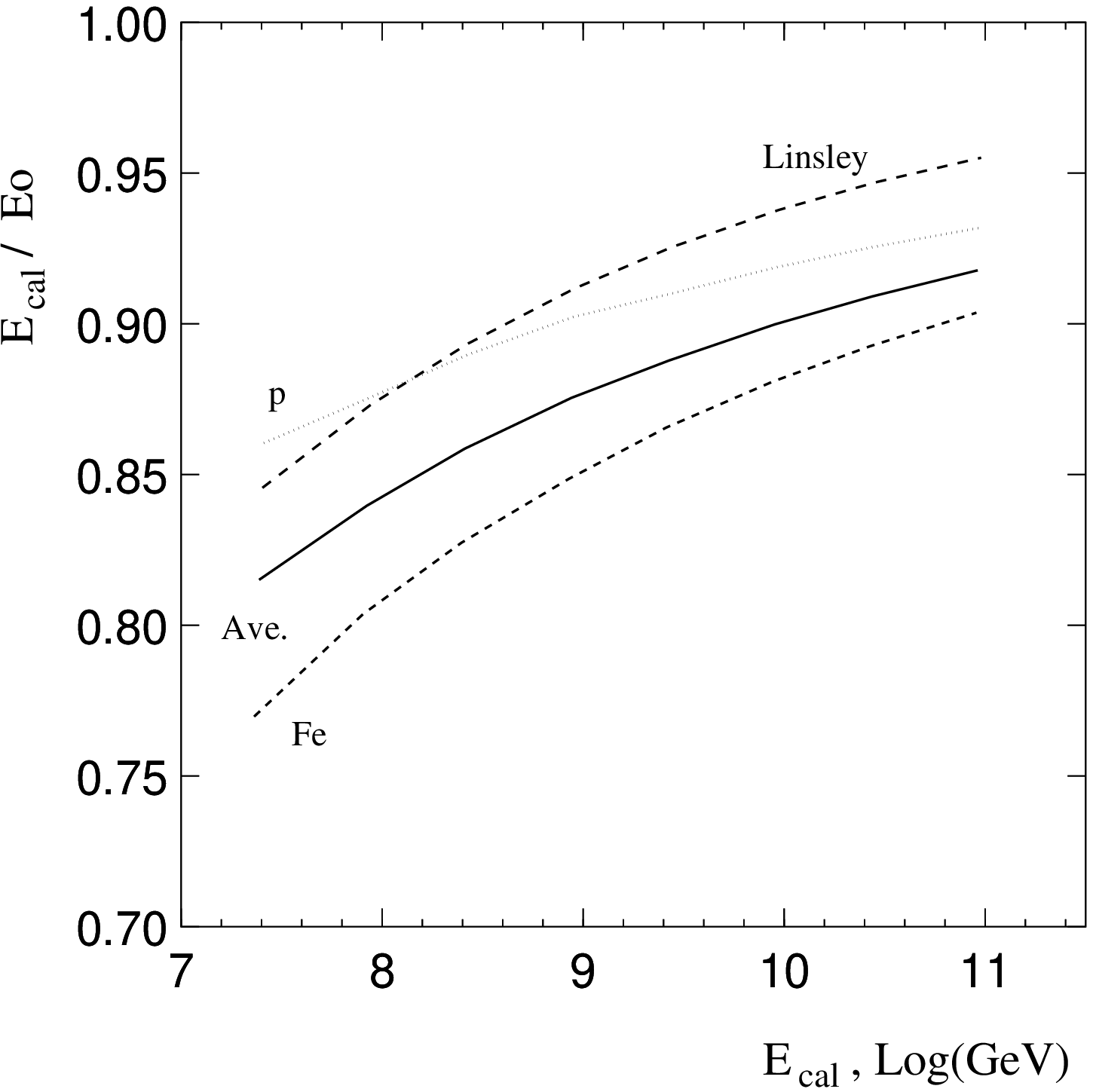}}
\end{center}
\caption{The functions for correcting the calorimetric energy to the
primary energy, as a function of calorimetric energy.  Shown are
the corrections for proton showers (dotted line) and iron
showers (short dashed line) and an average of the two(solid line). 
For comparison, Linsley's function is also shown.}
\label{fig.4}
\end{figure}


\vspace{1ex}



\begin{thebibliography}{000}

\bibitem{corsika} D. Heck, J. Knapp, J.N. Capdevielle, G. Schatz and 
T. Thouw, ``CORSIKA : A Monte Carlo Code to Simulate Extensive Air Showers'', 
Report FZKA 6019 (1998), Forschungszentrum Karlsruhe
\bibitem{rossi} B. Rossi, High Energy Particles, Ch. 5 (Prentice Hall, 1952)
\bibitem{egsg} A.F. Bielajew, ``Photon Monte Carlo simulation'',
Report PIRS-0393, National Research Council of Canada (1993)
(http://ehssun.lbl.gov/egs/epub.html)
\bibitem{egse} A.F. Bielajew and D.W.O. Rogers, ``Electron Monte Carlo 
simulation'', Report PIRS-0394, National Research Council of Canada (1993)
(http://ehssun.lbl.gov/egs/epub.html)
\bibitem{data} Particle Data Group, ``Review of Particle Physics'', 
The European Physical Journal, {\bf 3} (1998) 76, 148
\bibitem{table} M.J. Berger and S.M. Seltzer, ``Table of Energy Losses 
and Ranges of Electrons and Positrons'', National Aeronautics and Space
Administration Report NASA-SP-3012, (Washington DC, 1964)
\bibitem{linsley} J. Linsley, Proc. 18th ICRC, Bangalore, India, {\bf 12} 
(1983) 144
\bibitem{balt} R.M. Baltrusaitis {\it et al}., Proc. 19th ICRC, La Jolla, 
USA, {\bf 7} (1985) 159
\bibitem{fluo} F. Kakimoto {\it et al}., NIM {\bf A 372} (1996) 527
\bibitem{estar} M.J. Berger, ``ESTAR : Computer Program for Calculating 
Stopping Power'', NIST Report, NISTIR-4999, Washington DC (1992)
(http://physics.nist.gov/PhysRefData/Star/ Text/ESTAR.html)
\bibitem{qgsjet} N.N. Kalmykov, S.S. Ostapchenko and A.I. Pavlov, Nucl. Phys. B
(Proc. Suppl.) {\bf 52B} (1997) 17
\bibitem{radl} J. Linsley, Proc. 19th ICRC, La Jolla, USA, {\bf 7}
(1985) 193
\end{thebibliography}
\end{document}